\def\qquad{\hspace{30pt}}
\def\0{\newline}
\def\1{\newline}
\def\2{\par\noindent\newline}

\def\loq{,\kern-0.080em,\kern+0.05em}

\def\bfloq{{\bf,\kern-0.06em,\kern+0.05em}}

\def\footloq{,\kern-0.07em,\kern+0.03em}

\def\quabla{{\raise.7ex\hbox{\boxed{{}}}}}

\def\be{\begin{equation}}
\def\ee{\end{equation}}
\def\bea{\begin{eqnarray}}
\def\eea{\end{eqnarray}}
\def\bild#1#2#3{
\begin{figure}[ht]
\epsfxsize=#3cm
\begin{center}
\leavevmode
\epsffile{#1.eps}
\end{center}
\caption[]{#2}
\label{#1}
\end{figure}
}
 \documentstyle[12pt,epsf]{article}	       
\title{%
{\Large\bf
    \vspace*{-40pt}
Quantum ballistic transport
    \\[-07pt]                   
in in-plane-gate transistors showing
    \\[-09pt]                   
onset of a novel ferromagnetic phase transition%
\thanks{Presented at Third International Symposium on
        Nanostructures and Mesoscopic Systems,
        May 19-24, 1996, Santa Fe, New Mexico, USA}
}
     \\[15pt]  
      }
\author{%
{\sc Ralf D.\ Tscheuschner}
and
{\sc Andreas D.\ Wieck}
\\[15pt]  
Angewandte Festk\"orperphysik
\\[-07pt]
Ruhr-Universit\"at Bochum
\\[-07pt]
Universit\"atsstra\ss e 150
\\[-07pt]
D-44780 Bochum
\\[-07pt]
Federal Republic of Germany
\\[15pt]
       }
\date{{\it Received 21 May 1996}}
\begin{document}
\maketitle
\begin{abstract}
We study one-dimensional transport in focused-ion-beam written
in-plane-gate transistors on III-V heterostructures at moderately
low temperatures at {\it zero\/} bias {\it without any\/}
external magnetic field applied.
In accordance with a recent proposal of
{\sc A. Gold} and {\sc L. Calmels},
{\it Valley- and spin-occupancy instability in the quasi-one-dimensional
     electron gas\/}, Phil.\ Mag.\ Lett.\ {\bf 74}, 33-42 (1996)
and earlier experimental data, we observe plateaux in the source-drain
conductivity considered as a function of the gate voltage,
not only at multiples of $2 e^2/h$ but also clearly at $e^2/h$,
just before the channel closes to zero conductivity. This may be
interpreted as a many-electron effect, namely as a novel ballistic
ferromagnetic ground state evading standard descriptions and theorems.
\end{abstract}
%
\newpage		       
\pagestyle{myheadings}	       
\tableofcontents	       
\vfill\pagebreak	       
\newtheorem{A}{Axiom}	       
%
\noindent%
%
%
\section{Introduction}
In 1988 the quantization of the conductivity in one-dimensional short
channels was discovered [1]. Here we focus on the conductivity of the
lowest one-dimensional subbands, namely for an occupancy of one or two
of them. We find that the quantum of conductivity is
{\it not\/} $2e^2/h$ but simply $e^2/h$.
Our measurements are performed in
Al$_x$Ga$_{1-x}$As/GaAs-heterostructure
in-plane-gate (IPG) transistors operated
at moderately low temperatures ($1-2$\,K)
with {\it zero bias\/} and {\it zero external magnetic field\/}.
\par
We interpret our findings as a lifting of a spin occupancy degeneracy
within the spirit of a recent proposal by
{\sc Gold} and {\sc Camels} [2]
who claim that in one dimensional quantum
wires at low electron densities there
might exist a valley occupancy instability (in multivalley semiconductors
like Si) or rather a spin occupancy density (even in direct semiconductors
like GaAs).
\par
The response to an applied finite external magnetic field discloses the
intricate nature of the new state which seems to be a realization of
(longitudinally) polarized electron transport [3],
the polarization being of
spontaneous nature here. In fact, from a theoretical point of view it may be
regarded as some sort of ferromagnetic ground state evading the standard
spin model ({\sc Heisenberg}) picture
(and hence the associated low-dimensional no-go-theorems) [4]
rather being reminiscent of the itinerant
({\sc Bloch}) picture [5]
without being identical to the latter. In fact, in some
sense the observed state bears some resemblance to the notion of a chiral
(say gapless handed) fermion state stabilizing under a finite
${\bf E}\cdot{\bf B}$.
\par
In what follows we explain the principles of IPG devices, report the
experimental data, and give an outline of the theoretical perspective.
\section{IPG design and operation}
The principle of focused-ion-beam (FIB) implantation and the FIB-written
in-plane-gate (IPG) transistor are explained in
Figures 1-3.
\begin{enumerate}
\item
By direct implantation of an FIB line of approximately 50 nm width
we insulate two areas of the two-dimensional (2D) electron gas
against each other, that is, there does not flow any current (I) up to a
voltage (U) of some volts.
\item
By implantation of a somewhere disconnected FIB line there will
flow some current through the resulting narrow gap, forming a point
contact. The latter can be considered as a limit of a quasi-one-dimensional
(quasi-1D) current carrying channel.
\item
By implanting two FIB lines we form three regions defining a
\lq T{\rq}-shaped pattern leaving open a narrow gap at its vertex.
In this way source and drain regions are created.
Current flowing from source to
drain can be controlled by applying a voltage to the third region, the
gate. Thus we have a one-dimensional channel controlled
by a lateral field effect:
the IPG transistor [7].
\end{enumerate}
Note that the IPG transistor is an all-purpose device in the sense that it
works in the classical limit at room temperature as well as in the quantum
regime at low temperatures. In particular, our samples fabricated using IPG
technique are rugged enough to exhibit a very long lifetime, as compared,
for example, with more common {\sc Schottky} gate devices. At room
temperature the IPG transistor does work much like an ordinary junction
field effect transistor (JFET). In this work, however, we study only the
zero-bias (quantum) resistivity of the source-drain channel at moderately
low temperatures ($1-2$\,K).
%
%
\section{Experiments}
We grow a GaAs/Al$_{0.33}$Ga$_{0.67}$As modulation doped heterostructure by
molecular beam epitaxy on a semi-insulating GaAs (100) substrate. It
consists of a 2 $\mu$m nominally doped GaAs buffer layer and a 23 nm
undoped Al$_{0.33}$Ga$_{0.67}$As spacer layer, followed by 50 nm of Si-doped
Al$_{0.33}$Ga$_{0.67}$As and a 10 nm GaAs cap. The two-dimensional electron gas is
localized in a sheet within the GaAs buffer right at the interface to the
spacer.
The measured values for the electron sheet density and the electron
mobility are
\begin{itemize}
\item
at room temperature
$4.7 \times 10^{11}$\,cm$^{-2}$ and   6\,638 cm$^2$ V$^{-1}$ s$^{-1}$,
\item
at T = 77 K in the dark
$3.6 \times 10^{11}$\,cm$^{-2}$ and 166\,900 cm$^2$ V$^{-1}$ s$^{-1}$,
\item
at T =  5 K in the dark
$3.1 \times 10^{11}$\,cm$^{-2}$ and 666\,000 cm$^2$ V$^{-1}$ s$^{-1}$,
\end{itemize}
respectively. By \lq\lq in the dark\rq\rq\
we indicate that we measure all values
without illumination prior to or during
the {\sc Hall} experiment.
\par
After the growth process the sample is mesa-etched into a standard
{\sc Hall}-bar geometry with a width
of 150\,$\mu$m and a distance of 200\,$\mu$m between
ohmic contacts which were made with an AuGe/Ni alloy. Using an
focused-ion beam (FIB) of 100 keV Ga+ ions, nominally 50 nm wide insulating
lines were directly written at normal incidence with a dose of
$1 \times 10^{12}$\,cm$^{-2}$
to define in-plane gated channels of 1\,$\mu$m width.
No subsequent annealing
has been applied to the FIB-written line.
\par
Transport measurements are performed at T = $1-2$\,K in the dark with zero
source-drain bias without magnetic field applied. In addition, the sample
can be mounted in two different orientations with respect to an external
magnetic field, such that the latter can be applied perpendicularly to the
sample's surface or else in direction parallel to the conducting channel.
We measure the a.c.\ voltage drop between source and drain in injecting an 1
nA alternating c.w.\ current with constant r.m.s.\ value. The chosen
frequency is 81\,Hz and voltage detection is performed with a standard
lock-in setup including an {\sc Ithaco Dynatrac} 391A lock-in amplifier
supplemented by a {\sc Hewlett-Packard} 4142B parameter analyzer. All
measurements are done by first sweeping the gate voltage down to
negative values and then sweeping up again. This always results in double
traces in the figures, showing the reproducibility of the data.
\section{Data analysis}
At zero bias the measured nominal
source-drain voltage $V_{sd}$ devided by the
constant injection current $I_{sd}$ equals
the sum $R$ of the resistance $R_{true}$ of the
constriction (which is of interest here)
and the series resistance $R_s$ built of
the areas between the contacts.
The latter has to be subtracted in order to get
the correct values from which we can read off the quantized values.
Unfortunately we do not know its exact value. Denoting the i-th trial
subtraction resistance by $R_i$
we obtain a family (i = 1, 2, ...) of conductivity
curves of the form
\begin{equation}
\sigma_i(V_g)
=
\frac{1}{R_{true}(V_g)+R_s-R_i},
\end{equation}
from which we pick off the one which has the best overall linear rising
behaviour with increasing gate voltage $V_g$. For further details concerning
the resistance $R_s$ of the triangular-shaped entrance and outlet of the
channel the reader is referred to Ref.\ [7].
\par
Figures 4 and 5 show data taken during different sessions from
measurements without an external magnetic field applied. The curves with
the overall negative slope show the resistivities versus the gate voltage, the
curves with the overall positive slope show the conductivities versus the
gate voltage, respectively. The observed conductivity structure is
independent of sweep direction, i.e. there are no single-shot artifacts.
There is not only a quantization step close to
$2e^2/h$,
but also a clear plateau at
$e^2/h$
even though there is no magnetic field applied at all. The next
quantization occurs at
$4e^2/h$
which is consistent with theory, predicting the
effect at
$3e^2/h$
to be absent [2].
The structure at
$e^2/h$
can be observed in all
samples we prepare.
However, it is not always as pronounced.
Figures 6-9
show analogous curves in case of a perpendicular and
longitudinal external magnetic field. A magnetic field perpendicular to the
sample's surface suppresses the effect whereas a magnetic field parallel to
the direction of transport at least does not have any influence on its
appearance.
\section{Conclusion}
By using focused ion beams we define small structures within the plane of
a two-dimensional electron gas.
In particular, by controlling a quasi-1D
channel its electrical width is squeezed down to the quantum
wire regime. Though one must admit that ion beam damage adversely
influences the electron mobility in the channel it is expected that
quantization effects can be seen most clearly at the point just before the
channel closes. In fact, this is what has been observed.
\par
Our main finding is a conductivity quantization step at
$e^2/h$
in an in-plane-gate transistor operated
at moderately low temperatures ($1-2$\,K)
with {\it zero bias\/} and {\it zero external magnetic field\/}.
Some indications of this effect may
be extracted from the data in the work by
{\sc de\,Vries} {\it et al.\/} [8].
The analogous effect was not seen
in recent high-precision experients by
{\sc Thomas} {\it et al.\/}
using split gate devices [9].
However, the evolution of half plateaus as a
function of an additional d.c.\ electric field (and perpendicular external
magnetic field) in a ballistic quasi-1D constriction have been
found by
{\sc Patel} {\it et al.\/} [10].
Evidently, in-plane-gate devices differ from
split-gate-devices in one essential aspect, namely in the geometrical form
of the lateral band structure, for which we have no general theory so far.
We have to take into account that the physical situation in our devices is
not exactly equivalent to the one in split-gate setups. Deviations from the
exact quantization rule are most probably due to ionized impurity
scattering in the sample [11].
However, there is definitely an additional
structure giving us notice of new physics in ballistic electron transport.
\par
We interpret our findings
at {\it zero bias\/} and {\it zero external magnetic field\/} as a
true many-body effect in a one-dimensional electron system. More
speculatively, it may be a form of a novel ferromagnetical ordered state
which we christened \lq\lq ferromagnetism on the wing\rq\rq.
The calculations of {\sc Gold} and {\sc Camels} [2]
suggest the existence of a spin instability at low
electron density in a GaAs quantum wire. This mechanism should manifest
itself in exactly the same way as shown up in our experiment.
Consequently, it fits well into the framework of spin-polarized transport
[3],
the polarization here being due to an electron-electron correlation.
This explanation is adjacent because of the apparently spontaneous nature
of the state and its response to the external conditions which seems to
favour a stability under a finite ${\bf E}\cdot{\bf B}$.
Future studies should elaborate on this fact.
\section{Acknowledgements}
One of us (A.D.W.) would like to thank
{\sc A. Gold} for sending him a preprint
of his paper.
For R.D.T. it is a great pleasure to thank
(a) {\sc A. Fischer} for
    introducing him to the secrets of MBE growth and
(b) {\sc D.\ de\,Vries} for
    giving him some lessons
    on the FIB/IPG stuff during his stay
    at the Max-Planck-Institut f\"ur
    Festk\"orperforschung, Stuttgart, in 1995.
Finally he is also very grateful to
{\sc R.L.\ Stuller}, {\sc H. He\ss ling},
and Professor {\sc N.\ Schopohl}
for inspiring discussions.
\newpage 
\section{References}
\begin{itemize}
\item[{[1]}]
{\sc B.J.\ van Wees},
{\sc H.\ van Houten},
{\sc C.W.J.\ Beenakker},
{\sc J.G.\ Williamson},\linebreak
{\sc L.P.\ Kouwenhoven},
{\sc D.\ van der Marel},
and
{\sc C.T.\ Foxon},
{\it Quantized conductance of point contacts
    in two-dimensional electron gas\/},
Phys.\ Rev.\ Lett.\ {\bf 60}, 848-850 (1988);\\
{\sc D.A.\ Wharam},
{\sc T.J.\ Thornton},
{\sc R.\ Newbury},
{\sc M.\ Pepper},
{\sc H.\ Ahmed},
{\sc J.E.F.\ Frost},
{\sc D.G.\ Hasko},
{\sc D.C.\ Peacock},
{\sc D.A.\ Richie},
and
{\sc G.A.C.\ Jones},
{\it One-dimensional transport an the
     quantisation of the ballistic resistance\/},
J.\ Phys.\ {\bf C21}, L209-L214 (1988);\\
{\sc D.A.\ Wharam},
{\sc M.\ Pepper},
{\sc H.\ Ahmed},
{\sc J.E.F.\ Frost},
{\sc D.G.\ Hasko},
{\sc D.C.\ Peacock},
{\sc D.A.\ Richie},
and
{\sc G.A.C.\ Jones},
{\it Addition of the one-dimensional quantised ballistic resistance\/},
J.\ Phys.\ {\bf C21}, L887-L891 (1988);\\
{\sc D.A.\ Wharam},
{\sc R.\ Newbury},
{\sc M.\ Pepper},
{\sc D.G.\ Hasko},
{\sc H.\ Ahmed},
{\sc J.E.F.\ Frost},
{\sc D.A.\ Richie},
{\sc D.C.\ Peacock},
 {\sc G.A.C.\ Jones},
{\sc T.J.\ Thornton},
and
{\sc U.\ Ekenberg},
{\it Ballistic electron transport in
     quasi-one-dimensional systems\/},
Surf.\ Sci.\ {\bf 229}, 233-238 (1990).
\item[{[2]}]
{\sc A.\ Gold}
and
{\sc L.\ Calmels},
{\it Valley- and spin-occupancy instability in the
     quasi-one-dimensional electron gas\/},
Phil.\ Mag.\ Lett.\ {\bf 74}, 33-42 (1996).
\item[{[3]}]
{\sc G.A.\ Prinz},
{\it Spin-polarized transport\/},
Physics Today {\bf 4/95}, 58-63 (1995).
\item[{[4]}]
{\sc W.\ Heisenberg},
{\it Zur Theorie des Ferromagnetismus\/},
Z.\ Phys.\ {\bf 49}, 619-636 (1928);\\
{\sc N.D.\ Mermin}
and
{\sc H.\ Wagner},
{\it Absence of ferromagnetism or antiferromagnetism in
     one- or two-dimensional isotropic Heisenberg models\/},
Phys.\ Rev.\ Lett.\ {\bf 17}, 1133-1136 (1966).
\item[{[5]}]
{\sc F.\ Bloch},
{\it Bemerkung zur Elektronentheorie
     des Ferromagnetismus und der elektrischen Leitf\"ahigkeit\/},
Z.\ Phys.\ {\bf 57}, 545-555 (1929);\\
{\sc C.\ Herring},
{\it Exchange interactions among itinerant electrons\/}
in
{\it Magnetism Vol.\,IV\/},
ed.\ by
{\sc G.T.\ Rado and H.\ Suhl},
Academic Press, New York and London, 1966;\\
{\sc D.C.\ Mattis} (ed.),
{\it The many body problem\/},
World Scientific, Singapore, 1993.
\item[{[6]}]
{\sc A.D.\ Wieck}
and
{\sc K.\ Ploog},
{\it In-plane-gate quantum wire transistor
    fabricated with directly written focused ion beams\/},
Appl.\ Phys.\ Lett.\ {\bf 56}, 928-930 (1990);\\
{\sc A.D.\ Wieck}
and
{\sc K.\ Ploog},
{\it High transconductance in-plane-gated transistors\/},
Appl.\ Phys.\ Lett.\ {\bf 61}, 1048-1050 (1992).
\item[{[7]}]
{\sc D.K.\ de Vries}
and
{\sc A.D.\ Wieck},
{\it Series resistance of in-plane-gated transistors
     and quantum point contacts\/},
J.\ Vac.\ Sci.\ Technol.\ {\bf B13}, 394-395 (1995).
\item[{[8]}]
{\sc D.K.\ de Vries},
{\sc K.\ Ploog},
and
{\sc A.D.\ Wieck},
{\it Quasi-one-dimensional ballistic electron transport
     in in-plane-gate channels at liquid-nitrogen temperature\/},
Solid State Electronics {\bf 37}, 701-703 (1994).
\item[{[9]}]
{\sc K.J.\ Thomas},
{\sc M.Y.\ Simmons},
{\sc J.T.\ Nicholls},
{\sc D.R.\ Mace},
{\sc M.\ Pepper},
and
{\sc D.A. Ritchie},
{\it Ballistic transport in one-dimensional constrictions
     formed in deep two-dimensional electron gases\/},
Appl.\ Phys.\ Lett.\ {\bf 67}, 109-111 (1995).
\item[{[10]}]
{\sc N.K.\ Patel},
{\sc L.\ Mart\'in-Moreno},
{\sc M.\ Pepper},
{\sc R.\ Newbury},
{\sc J.E.F.\ Frost},
{\sc D.A.\ Ritchie},
{\sc G.A.C.\ Jones},
{\sc J.T.M.B.\ Janssen},
{\sc J.\ Singleton},
and
{\sc J.A.A.J.\ Perenboom},
{\it Ballistic transport in one dimension:
     additional quantisation produced by an electric field\/},
J.\ Phys.\ {\bf C2}, 7247-7254 (1990);\\
{\sc N.K.\ Patel},
{\sc J.T.\ Nicholls},
{\sc L.\ Mart\'in-Moreno},
{\sc M.\ Pepper},
{\sc J.E.F.\ Frost},
{\sc D.A.\ Ritchie},
and
{\sc G.A.C.\ Jones},
{\it Evolution of half plateaus as a function of electric field
     in a ballistic quasi-one-dimensional constriction\/},
Phys.\ Rev.\ {\bf B44}, 13549-13555 (1991).
\item[{[11]}]
{\sc G.\ Timp},
{\it When does a wire become an electron waveguide?\/}
Semiconductors and Semimetals {\bf 35}, 113-190 (1992);\\
{\sc E.G.\ Haanappel}
and
{\sc D.\ van der Marel},
{\it Conductance oscillations in two-dimensional
     Sharvin point contacts\/},
Phys.\ Rev.\ {\bf B39}, 5484-5487 (1989);\\
{\sc D.\ van der Marel}
and
{\sc E.G.\ Haanappel},
{\it Model calculations of the quantum ballistic
     transport in two-dimensional constriction-type
     microstructures\/},
Phys.\ Rev.\ {\bf B39}, 7811-7820 (1989);\\
{\sc C.S.\ Chu}
and
{\sc R.S.\ Sorbello},
{\it Effects of impurities on the
     quantized conductance of narrow channels\/},
Phys.\ Rev.\ {\bf B40}, 5941-5949 (1989).
\end{itemize}
%
%
\newpage
\section{Figures}
\vspace*{2cm}
\bild{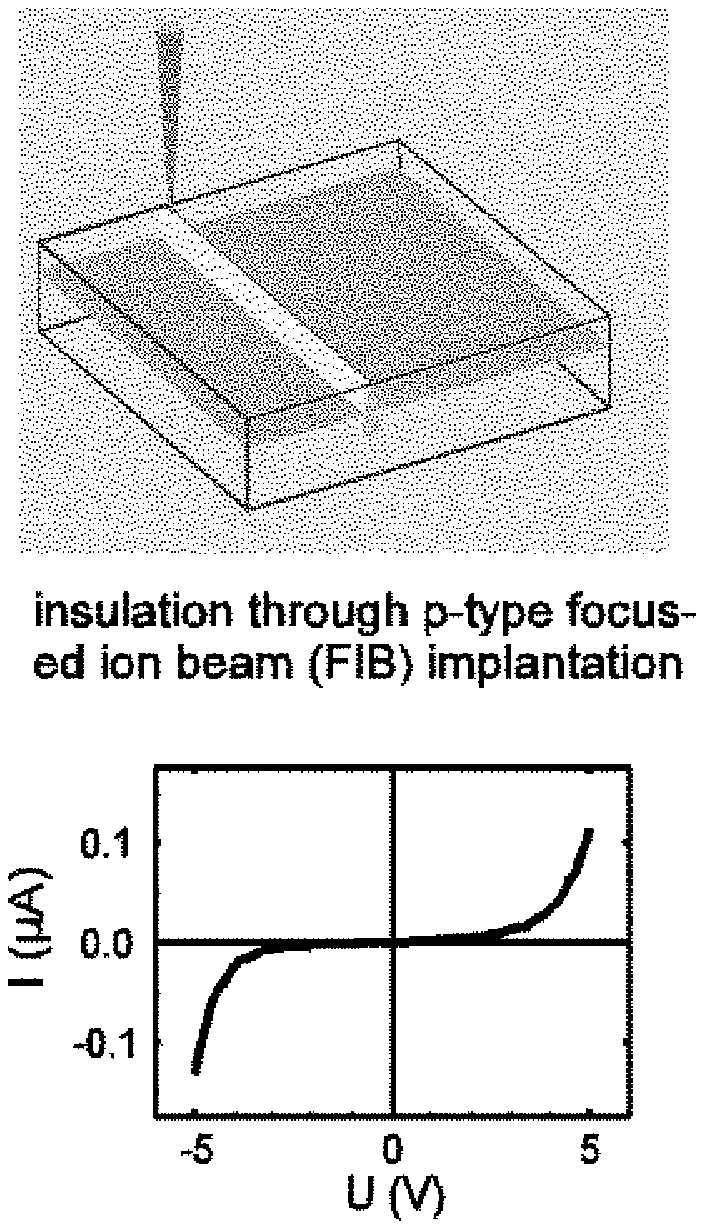}
     {Evolution of insulating lines in a two-dimensional electron gas
      to an in-plane-gate (IPG) transistor (Part 1 of 3).}
     {9}
\bild{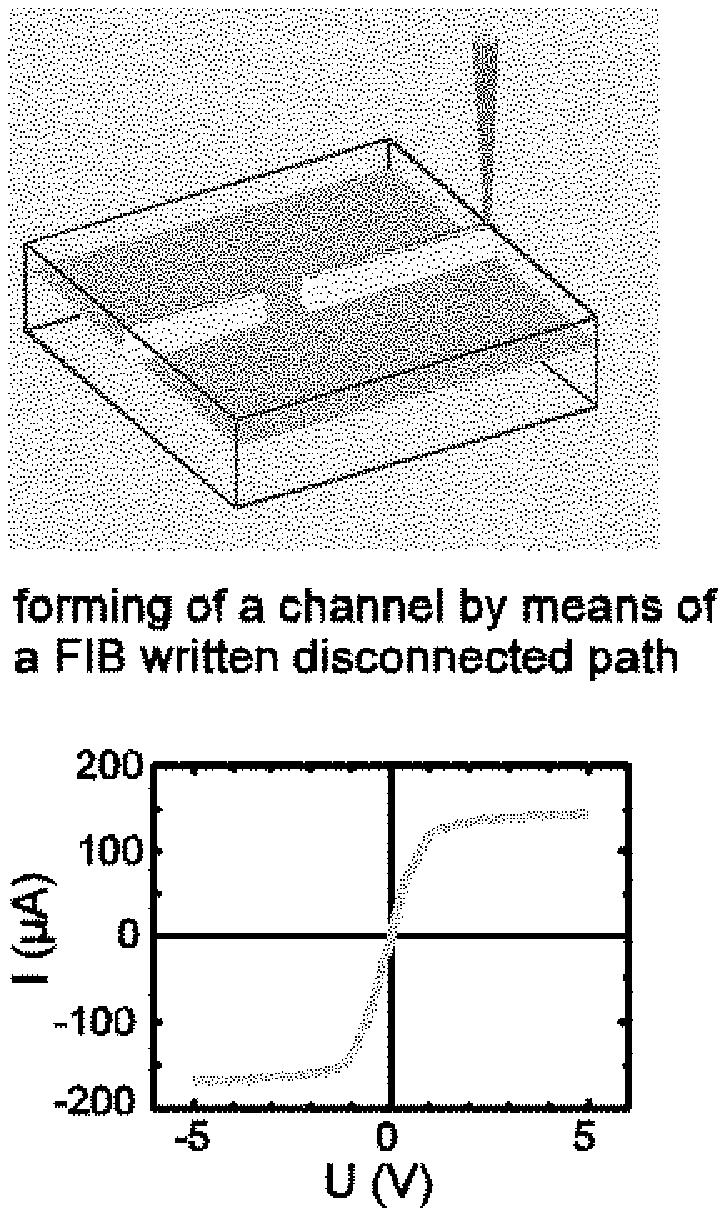}
     {Evolution of insulating lines in a two-dimensional electron gas
      to an in-plane-gate (IPG) transistor (Part 2 of 3).}
     {9}
\bild{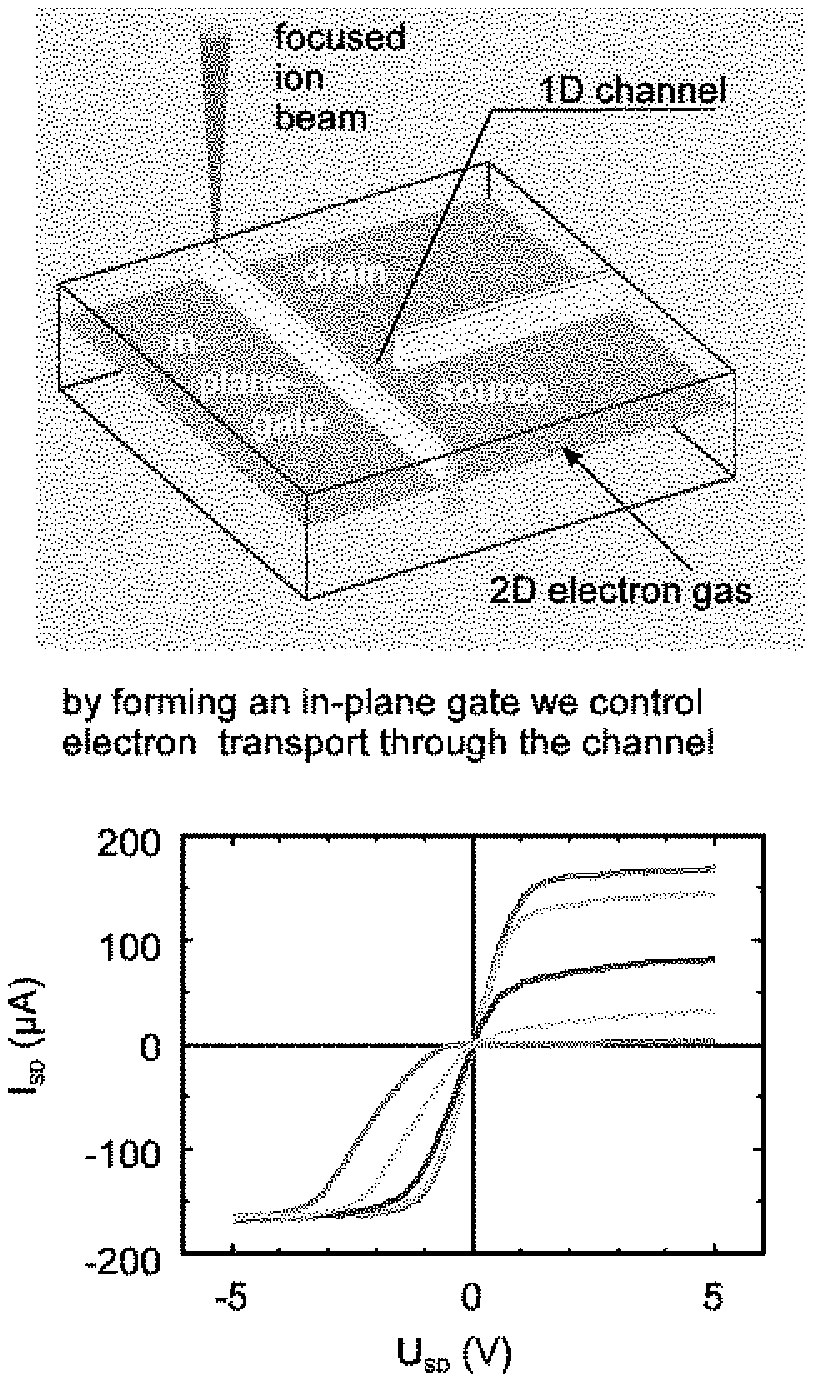}
     {Evolution of insulating lines in a two-dimensional electron gas
      to an in-plane-gate (IPG) transistor (Part 3 of 3).}
     {9}
\bild{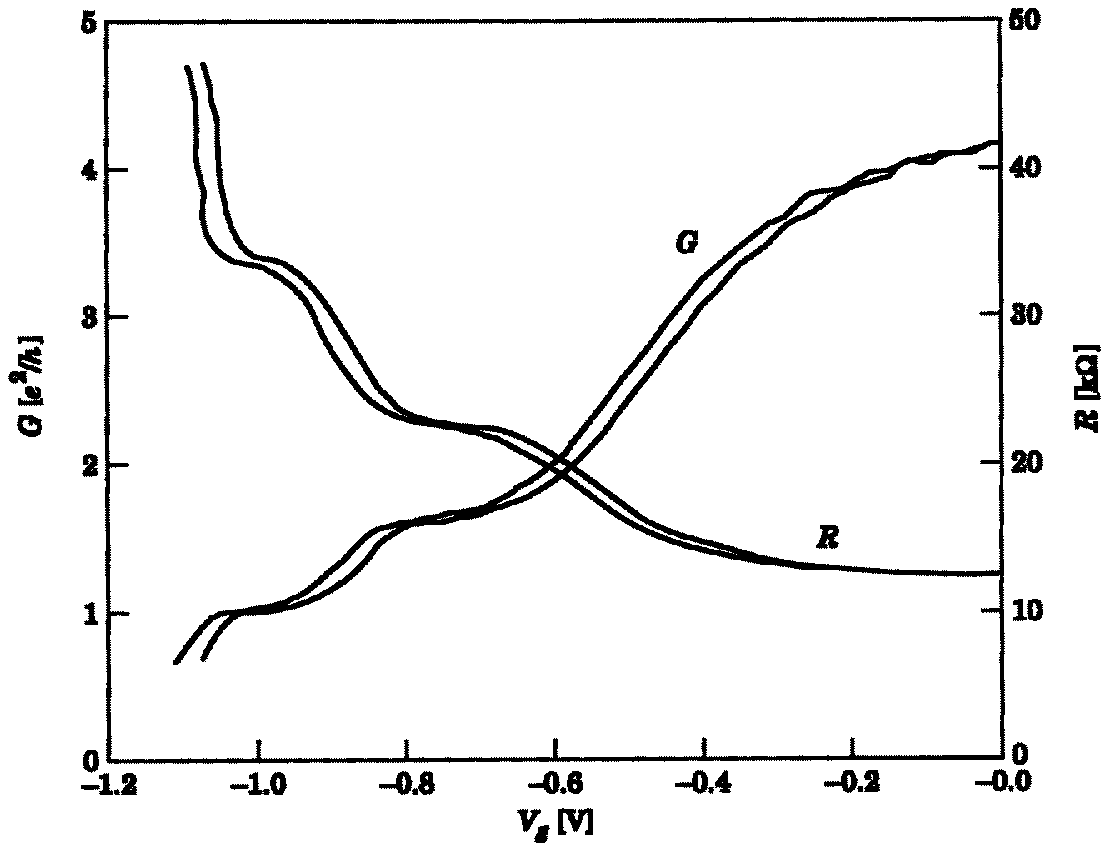}
     {Zero-bias conductivity $G$
      (in units of $e^2/h$)
      and resistance R (in k$\Omega$)
      of an IPG transistor at low temperatures
      T = 1.5\,K versus the IPG voltage $V_g$ (in V) at B = 0.}
     {13}
\bild{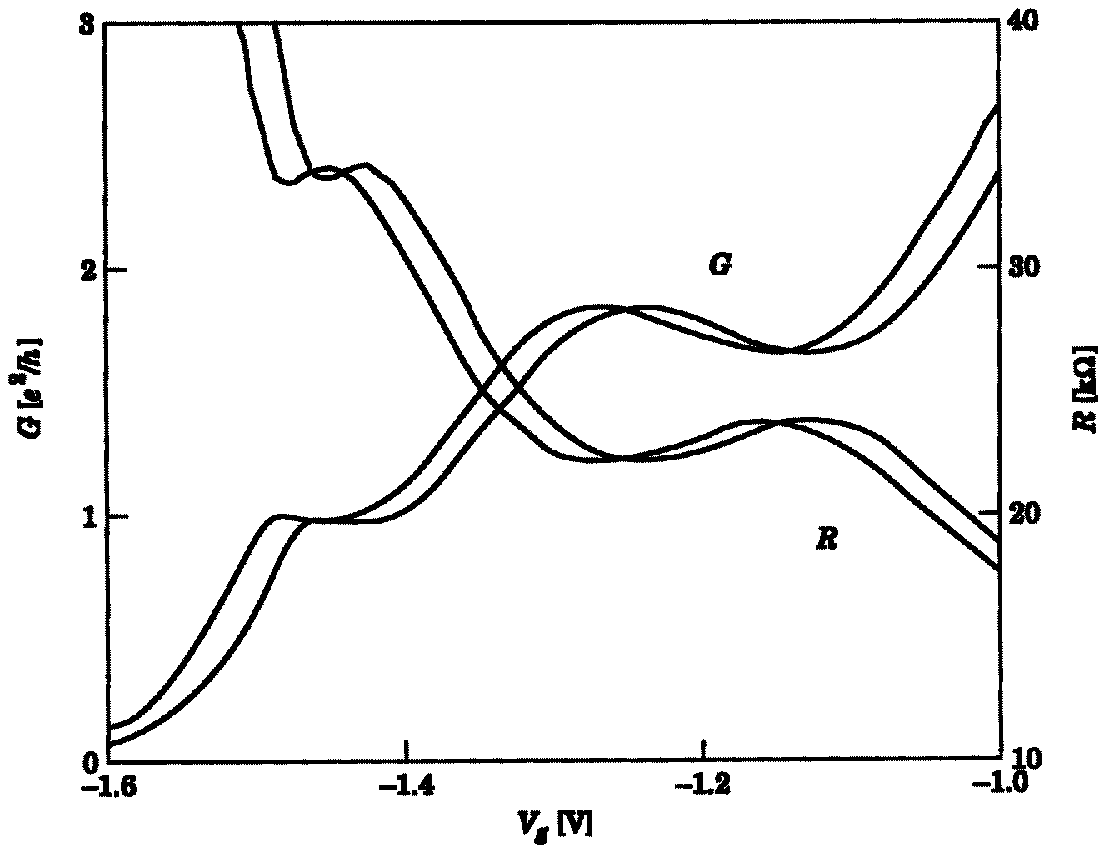}
     {Same as Fig.\ 4,
      but with different scaling factors.
      These data are
      taken during a later measurement.}
     {13}
\bild{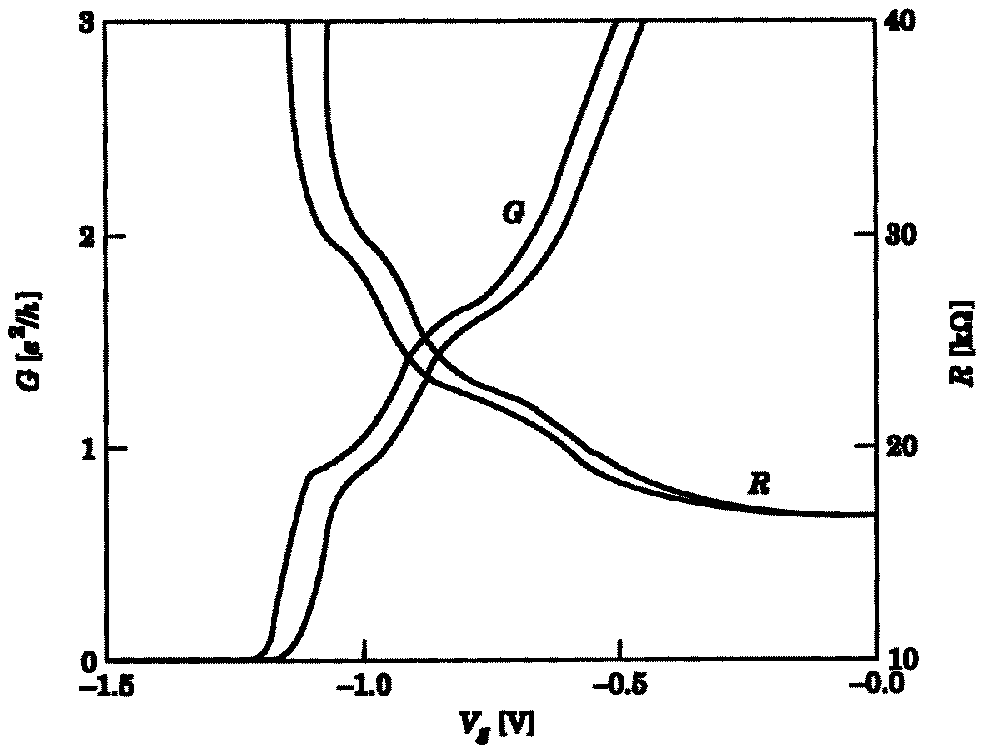}
     {Same as Fig.\ 4 and 5.
      Note that for gate voltages
      $V_g<-1.2$\,V
      the resistance increases monotonically,
      i.e. there is no conductivity plateau below
      $e^2/h$.}
     {13}
\bild{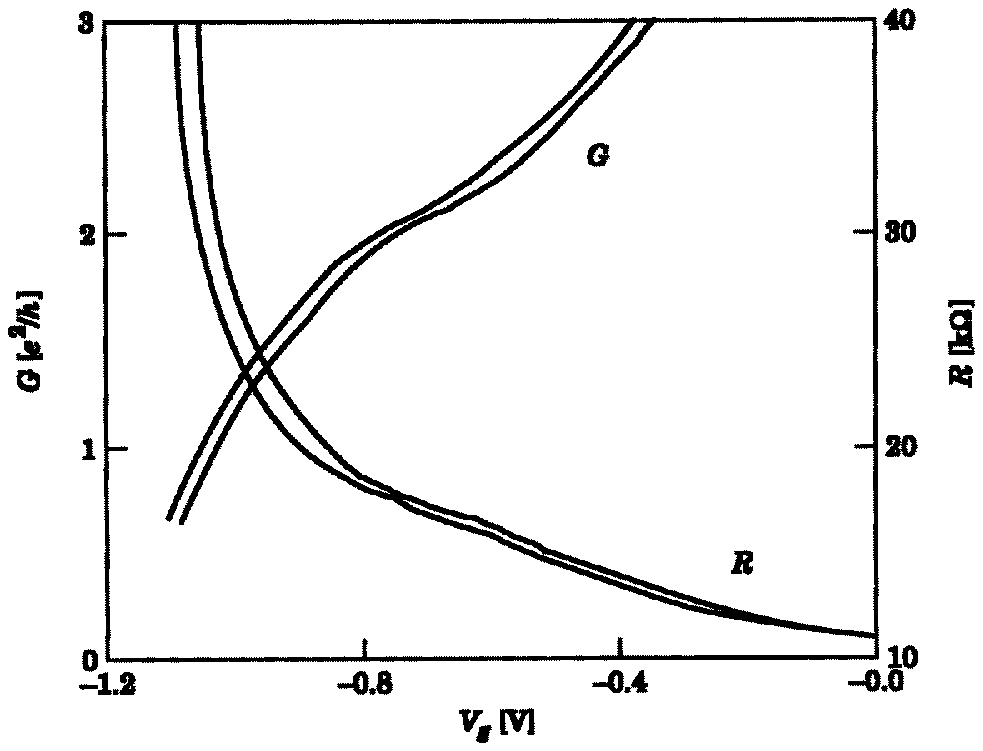}
     {The same measurement as in Fig.\ 6,
      but with a magnetic field applied perpendicular to the
      sample's surface, giving a finite
      ${\bf E}\times{\bf B}$ value.
      Quantizations are suppressed.}
     {13}
\bild{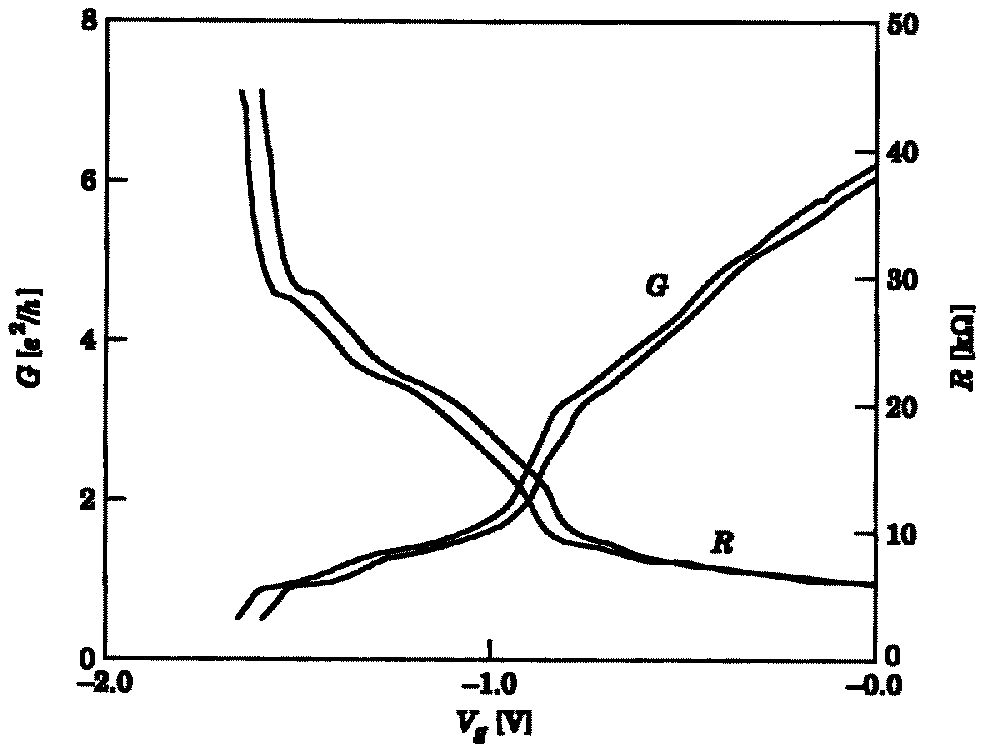}
     {Same measurement as in Figs.\ 6-7
      without a magnetic field.}
     {13}
\bild{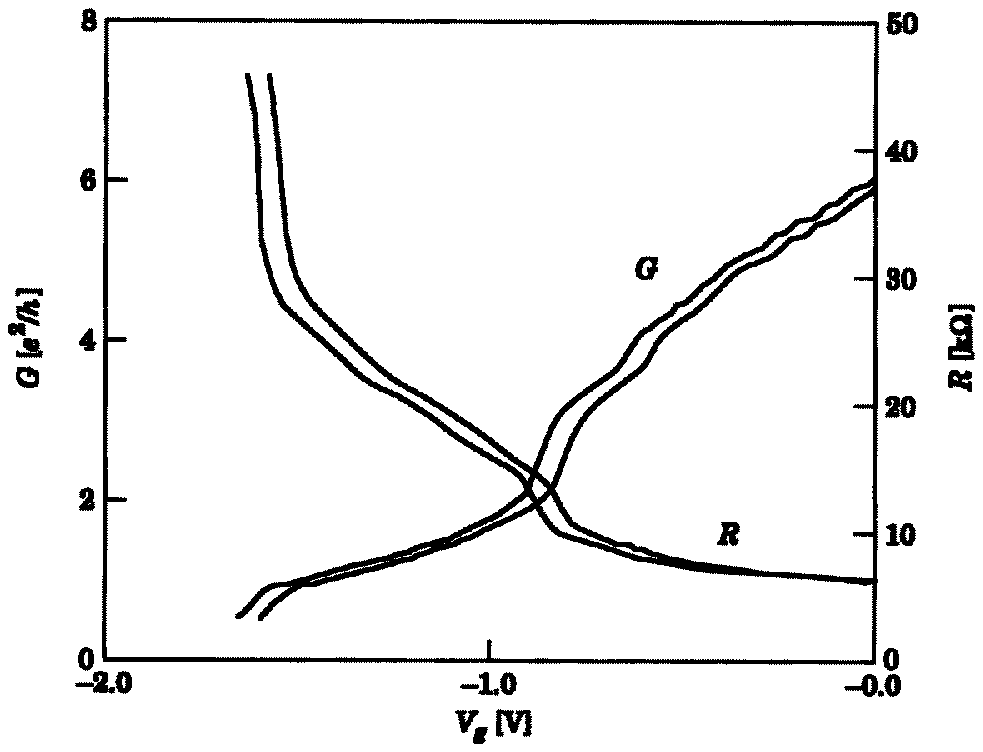}
     {Same measurement as in Fig.\ 6-7,
      but with B = 3\,T, applied in transport direction
      of the 1D channel giving a finite
      ${\bf E}\cdot{\bf B}$ value.
      All ballistic features are conserved.}
     {13}
\end{document}